\documentclass[cameraready]{Interspeech}

\usepackage{multirow}
\usepackage{float}
\usepackage{booktabs}
\usepackage{comment}
\usepackage{makecell}
\usepackage{url}

\usepackage[T1]{fontenc}
\usepackage[utf8]{inputenc} 
\usepackage{tipa}

\title{Using Phonological-Level Wav2Vec2 for Mandarin Automatic Mispronunciation Detection and Diagnosis}

\author[affiliation={1}, orcid=0009-0004-3707-0747]{Jinghao}{Chen}
\author[affiliation={1}, orcid=0000-0002-1091-8531]{Mostafa}{Shahin}
\author[affiliation={1}, orcid=0000-0002-1240-6572]{Beena}{Ahmed}

\address
{
    $^1$ School of Electrical Engineering and Telecommunications, UNSW, Australia
}
\email{z5327748@unsw.edu.au, m.shahin@unsw.edu.au, beena.ahmed@unsw.edu.au}

\begin{document}
\maketitle
\begin{abstract} 
Automatic mispronunciation detection and diagnosis (MDD) plays a crucial role in L2 Mandarin pronunciation learning. While end-to-end (E2E) based MDD methods have substantially improved phoneme-level detection accuracy, diagnostic feedback remains limited, as segmental and tonal errors are not explicitly separated. In this paper, we propose a phonological feature-based MDD framework that models both segmental and tonal attributes within a unified Wav2Vec2-CTC architecture. Experimental results show that the proposed method reduces False Acceptance Rate (FAR) by 10.1\% and Diagnostic Error Rate (DER) by 23.6\% compared with the phoneme-only baseline system. By decomposing phonemes into low-level phonological components, the proposed approach enables more detailed and interpretable diagnostic feedback for L2 learners.
\end{abstract}

\section{Introduction}

Effective and informative MDD remain major challenges in Computer-Aided Pronunciation Learning (CAPL). Mandarin presents unique challenges for MDD due to its tightly integrated segmental–tonal phonological system and rich contrastive phoneme inventory \cite{CRNN1, MandarinPhono}. Several  Mandarin learning systems integrate phoneme-level mispronunciation detection (MD) and provide visual articulatory feedback to learners \cite{MandarinApp}. However, the feedback provided often does not include details of how the errors are generated during the process.

Mandarin syllables consist of closely coupled initial–final structures with lexical tone realised through distinct pitch contours~\cite{Hanyu, fourTone}. Therefore, accurate MDD requires a phonetic representation that consistently captures both segmental articulation and tonal realisation. In Mandarin Learning, Hanyu Pinyin~\cite{IPAandPinyin} is widely used. But it does not explicitly distinguish certain context-dependent vowel pronunciations~\cite{Allophone}, requiring additional processing to recover these distinctions for analysis. In contrast, the International Phonetic Alphabet (IPA)~\cite{IPABOOK} preserves these distinctions through separate phonetic symbols.

Early automatic speech recognition (ASR) approaches to Mandarin MD relied on Goodness of Pronunciation (GOP)-based posterior scoring \cite{DNN1, DNN3}, followed by sequence-aware models such as Connectionist Temporal Classification (CTC)-trained Convolutional Recurrent Neural Networks to better capture temporal patterns \cite{CRNN1}. 
More recently, E2E systems leveraging self-supervised representations (e.g., wav2vec2) 
and advanced encoders (e.g., Conformer) have substantially improved detection performance \cite{e2e1,e2e2, conform}, with both CTC and transducer-based paradigms explored 
for pronunciation modelling and adaptation \cite{BIDFSM, RNN-T, Meta}. However, these advances mainly target detection accuracy. Diagnostic feedback is typically reduced to scores or phoneme-level judgments, without explicitly modelling segmental and tonal attributes, leaving detailed and informative diagnosis underdeveloped.

Speech-attribute-based modelling has been explored to provide more interpretable diagnostic feedback by decomposing phonemes into articulatory and phonological components. Recent studies have demonstrated its effectiveness in non-native English pronunciation assessment, including DNN-GOP-based models \cite{li2016} and self-supervised phonological approaches built upon wav2vec2 \cite{Mostafa}. These works show that phonological attributes can serve as informative intermediate representations for fine-grained analysis of pronunciation.

While attribute-based modelling has demonstrated effectiveness in non-tonal languages, its adaptation to Mandarin remains underexplored. Existing studies on Mandarin speech attributes are relatively limited and mainly focus on articulatory modelling for ASR robustness \cite{att2007,att2011}. Moreover, Mandarin introduces additional challenges due to its tonal nature. Although tone can be represented using pitch-related attributes that capture level and contour patterns \cite{2013pitch}, prior work has not integrated tone-related features within a unified phonological modelling framework. Consequently, segmental and tonal errors are typically analysed separately or represented using coarse tone labels, limiting detailed diagnostic capability.

In this work, we propose a phonological MDD framework for Mandarin. Our contributions are:
(1) a refined Mandarin speech-attribute inventory integrating segmental and tonal distinctions; 
(2) an attribute-level modelling framework built on a self-supervised backbone for joint segmental–tonal learning; 
(3) a multi-level diagnosis method linking phoneme-level confusions with attribute-level error patterns to provide interpretable pronunciation feedback. 
To our knowledge, this is the first systematic framework jointly modelling segmental and tonal attributes for Mandarin MDD. Code and scripts are available at \url{https://github.com/Evanchan1923/MDD_SpeechAttribute}.

\section{Methodology}
\begin{figure}[t]
  \centering
  \includegraphics[width=\linewidth]{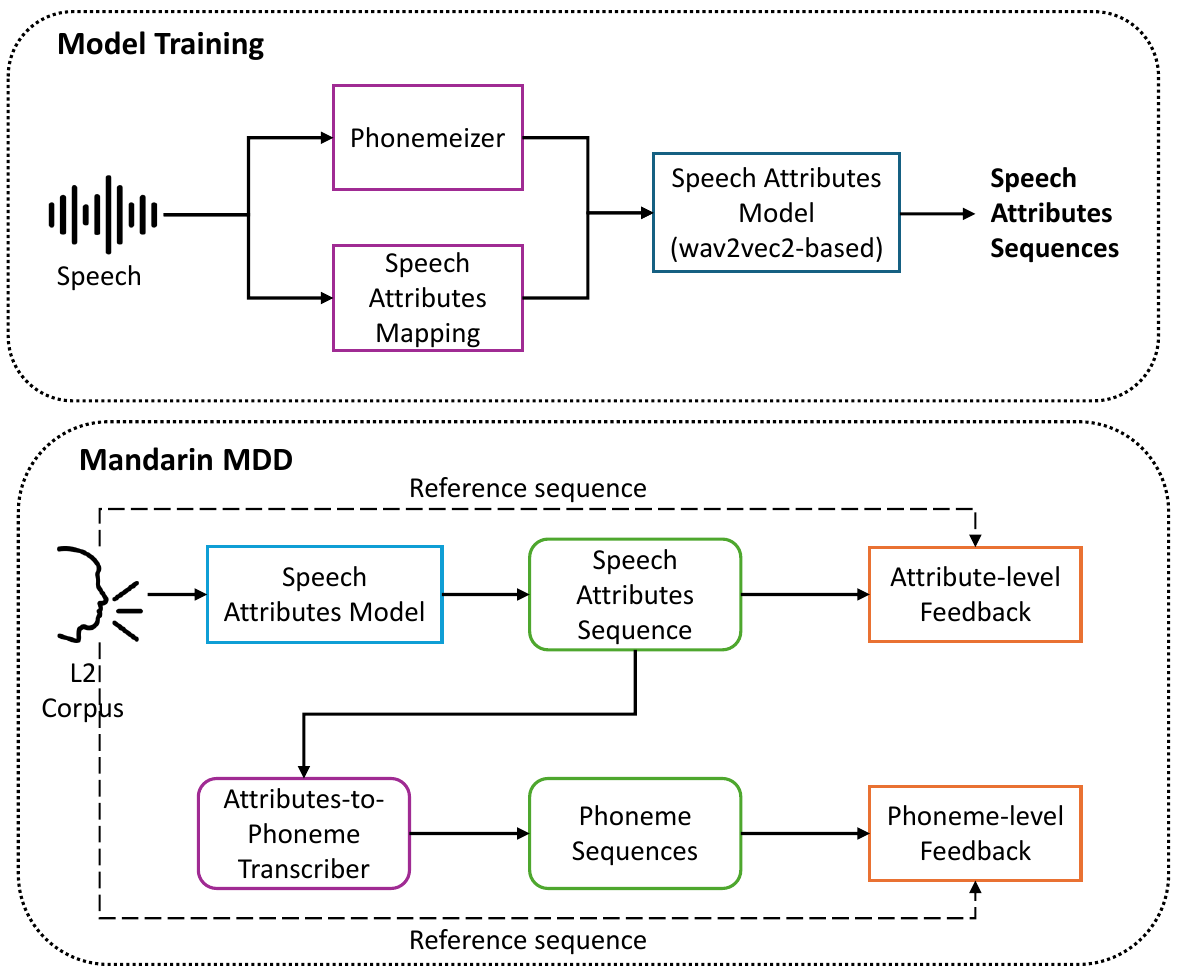}
  \caption{Overview of the proposed Mandarin phonological MDD framework.}
  \label{fig:framework}
\end{figure}

Fig.~\ref{fig:framework} illustrates the proposed Mandarin-oriented phonological MDD framework, consisting of two stages: (i) training a speech-attribute model on native Mandarin speech and (ii) applying it to L2 speech for mispronunciation diagnosis.

During training, phoneme sequences from transcripts are mapped to binary phonological attributes using a predefined phoneme-to-attribute mapping. A wav2vec 2-based CTC model is then fine-tuned to predict attribute sequences from speech using a multi-label objective \cite{Mostafa}, in which each attribute is represented as a binary (+att / -att) sequence. At inference, the model predicts attribute sequences for L2 speech. Attribute-level feedback is obtained by comparing predicted and reference attributes, while phoneme-level feedback is derived by converting predicted attributes into phoneme sequences using an attributes-to-phoneme transcriber.

\subsection{Phonemizer}
We implement a Mandarin phonemization pipeline that converts transcripts into aligned Pinyin and IPA sequences~\cite{IPAandPinyin}. In the proposed system, Pinyin is used only as an intermediate romanisation layer for transcript conversion, alignment, and human-readable inspection; it is not used as the modelling representation. Since Pinyin does not encode certain context-dependent allophonic alternations~\cite{Allophone}, IPA is used for phoneme-level modelling and subsequent speech-attribute mapping. For instance, the Pinyin vowel /e/ may correspond to either /\textipa{e}/ or /\textipa{E}/ in IPA.

In practice, utterances are converted to space-delimited IPA sequences using Dragonmapper~\cite{Dragon}, with tones represented as numeric markers. In addition, multi-character phonemes (e.g., affricates and aspirated stops) are resolved using a curated longest-match strategy. Phoneme normalisation is later applied to the output following \textit{Xiandai Hanyu}~\cite{Hanyu} to ensure inventory consistency for phoneme variants. Consequently, this process yields a fixed IPA inventory used for subsequent attribute mapping. For example, the Mandarin word /ba1/, /dun4/ is phonemeised as /\textipa{p}/ /\textipa{a}1/ , /\textipa{t}/ /\textipa{w@n}4/.

\subsection{Speech Attribute Mapping}
The Speech-attribute mapping converts each IPA phoneme sequence into a structured binary representation for phonological MDD. Each phoneme is mapped to a fixed-dimensional vector organised into articulatory groups, including manner and place, vowel height, backness, rounding, diphthong, and tone, as summarised in Tables~\ref{tab:attr_inventory} and~\ref{tab:tonemap}. This representation decomposes phonological contrasts into interpretable components tailored to Mandarin phonology and serves as the supervision target for the MDD model.

Mandarin complex vocalic nuclei can be analysed either as single units or as 
sequences of vocalic targets~\cite{Hanyu, Diphthong}. Accordingly, we consider 
two IPA-based mappings: \textbf{IPA-S}, which keeps diphthongal or multivowel 
rimes as single units with dedicated vowel attributes, and \textbf{IPA-D}, which decomposes them into constituent monophthongs and removes multivowel-specific attributes. For example, /\textipa{iAU}/ is treated as one complex vocalic unit in IPA-S but decomposed into /\textipa{i}/, /\textipa{A}/, and /\textipa{U}/ in IPA-D.

Two tone modelling strategies are explored: \textbf{Tone-Cat}, which uses categorical tone labels, and \textbf{Tone-PT}, which represents tones using pitch-target descriptors~\cite{2024tonevalue}. These choices yield a $2\times2$ experimental setup. The resulting attribute inventories contain 44 and 50 dimensions for IPA-S under Tone-Cat and Tone-PT, respectively, and 40 and 46 dimensions for IPA-D.

\subsection{Feedback Generation}
Feedback is generated at both the attribute and phoneme levels after establishing token-level correspondence between predicted and reference sequences. To handle insertion and deletion errors (e.g., extra or missing segments), we apply Levenshtein alignment~\cite{Leven} between the predicted phoneme sequence and both the canonical and learner-realised transcriptions.

Based on the resulting alignment, phoneme-level feedback identifies substitutions, insertions, and deletions, and verifies whether detected mispronunciations match the learner’s realised output. Attribute-level feedback is then derived from aligned phoneme pairs by comparing the canonical attribute representation with the predicted attribute vector, highlighting missing, altered, or additional articulatory or tonal features as diagnostic cues. In addition, confusion pairs derived from the aligned phoneme and attribute sequences (e.g., target/predicted) are recorded to analyse systematic error patterns at both levels.

\begin{table}[ht]
  \caption{Mandarin phonological attribute inventory.}
  \label{tab:attr_inventory}
  \centering
  \footnotesize
  \setlength{\tabcolsep}{6pt}
  \renewcommand{\arraystretch}{1.1}
  \begin{tabular}{p{1.4cm} p{5.6cm}}
    \toprule
    \textbf{Category} & \textbf{Attributes} \\
    \midrule
    \textbf{Manner} 
      & plosive, nasal, fricative, affricate, lateral, approximant, 
       aspirated / unaspirated; voiced / voiceless \\
    \textbf{Place}  
      & bilabial, dentolabial, dental, alveolar, retroflex, 
       alveolopalatal, velar, labial--velar, labial--palatal \\
    \textbf{Vowel}         
      & height (close–open), backness (front–back), \\
      & rounding, diphthong type (rise / fall / center) \\
    \bottomrule
  \end{tabular}

\vspace{10pt}

  \caption{Tone attribute mapping: categorical labels (Tone-Cat) and pitch-target descriptors (Tone-PT).}
  \label{tab:tonemap}
  \centering
  \footnotesize
  \setlength{\tabcolsep}{8pt}
  \renewcommand{\arraystretch}{1.05}
  \begin{tabular}{c c c}
    \toprule
    \textbf{Tone} & \textbf{Tone-CAT} & \textbf{Tone-PT} \\
    \midrule
    Tone 1 & High      & Onset=5,\ Mid=5,\ Offset=5 \\
    Tone 2 & Rising    & Onset=3,\ Mid=4,\ Offset=5 \\
    Tone 3 & Dipping   & Onset=2,\ Mid=1,\ Offset=4 \\
    Tone 4 & Falling   & Onset=5,\ Mid=3,\ Offset=1 \\
    \bottomrule
  \end{tabular}
\end{table}
\section{Experiments and Results}

\subsection{Data}
This study uses three Mandarin speech corpora with distinct roles. We adopt the official train/validation/test split of Common Voice 13 (CV13-CN)~\cite{cv13} for model training. To evaluate cross-corpus generalisation, AISHELL-1 (AS-1)~\cite{aishell} is used exclusively for evaluation and is not involved in training. LATIC (LAT)~\cite{latic}, a non-native learner corpus annotated by proficient Mandarin speakers, is used only for MDD evaluation and is not included in training.

\begin{table}[ht]
  \caption{Overview of the Mandarin datasets used in this study.}
  \label{tab:datasets}
  \centering
  \scriptsize
  \setlength{\tabcolsep}{4pt}
  \begin{tabular}{c c c c}
    \toprule
    \textbf{Corpus} & \textbf{L1} & \textbf{Hours} & \textbf{Speakers} \\
    \midrule
    CommonVoice 13-CN & Mandarin & 67.2 & 288 \\
    AISHELL-1         & Mandarin & 170  & 400 \\
    LATIC             & \makecell{Russian, Korean,\\ French, Arabic} & 4 & 4 \\
    \bottomrule
  \end{tabular}
\end{table}

\subsection{Experimental Setup}
We adopt the pre-trained Wav2Vec 2.0 XLSR-53~\cite {w2v2} as the shared acoustic backbone. Speech-attribute supervision (Section~2.2) is applied for attribute-based systems. We also train a phoneme-level model as our baseline, using the same architecture and settings, without attribute modelling. Audio is resampled to 16 kHz and normalised, and non-Mandarin symbols are removed from transcripts. The CNN encoder is frozen during fine-tuning. Models are trained for 15 epochs using AdamW~\cite{adamW} with a peak learning rate of $5\times10^{-4}$ and a 15\% warm-up ratio. 
Gradient clipping (5.0) and ctc\_zero\_infinity are applied for stability.

\subsection{Training Configurations}
To analyse the effects of segmental and suprasegmental modelling, we conduct a 2×2 factorial study over two segmental mappings (IPA-S vs. IPA-D) and two tone representations (Tone-CAT vs. Tone-PT), as defined in Section~3.2. The combination of these factors yields four joint configurations (IPA-S/IPA-D × Tone-CAT/Tone-PT), each trained independently using the same backbone architecture and optimisation settings. In addition, three training settings are evaluated: Phoneme-only, Tone-only, and Phoneme+Tone, corresponding to segmental, tonal, and joint attribute prediction, respectively.

\subsection{Evaluation Metrics}
We evaluate speech-attribute recognition, phoneme recognition, and Mandarin MDD at both phoneme and attribute levels. For recognition tasks, Attribute Error Rate (AER) and Phoneme Error Rate (PER) are computed using Levenshtein~\cite{Leven} alignment, counting substitutions (S), insertions (I), and deletions (D) relative to the reference sequence. 

For MDD, each phoneme or attribute is categorised as True Acceptance (TA), False Rejection (FR), True Rejection (TR), or False Acceptance (FA) based on the canonical transcription, learner-realised transcription, and model prediction. TR cases are further divided into Correct Diagnosis (CD), when the prediction matches the learner-realised transcription, and Diagnosis Error (DE) otherwise. From these counts, we compute the False Acceptance Rate (FAR), False Rejection Rate (FRR), and Diagnosis Error Rate (DER), which measure undetected mispronunciations, incorrect rejections of correct pronunciations, and incorrect diagnoses, respectively.

\subsection{Cross-Corpus Attribute Recognition Performance}

In the phoneme-only experiments, IPA-D consistently outperforms IPA-S, reducing segmental AER from 3.38\% to 1.83\%, confirming that diphthong decomposition improves vowel modelling. In tone-only experiments, Tone-PT performs comparably to Tone-CAT with negligible differences in overall tone AER. However, attribute-level analysis reveals that the \textit{offset-5} attribute (associated with Tone 1 and 2) has the highest AER (8.35\%), suggesting difficulty in modelling high-level pitch.

Table~\ref{tab:cv13_aishell_aer_breakdown} presents AER under joint phoneme–tone modelling across different attribute mappings. The results show that diphthong decoupling remains the primary contributor to performance gains, with IPA-D achieving over 40\% relative reduction in AER compared with IPA-S. Within the IPA-D setting, Tone-CAT yields the lowest overall AER, whereas Tone-PT achieves the best tone-specific performance. This improvement in tonal modelling does not lead to the lowest overall AER because segmental components constitute the structural foundation of Mandarin syllables, making overall AER more sensitive to segmental modelling than to tonal refinement.

\begin{table}[ht]
\centering
\footnotesize
\caption{Cross-corpus AER ($\downarrow$) on AISHELL-1.
Tone, consonant (Cons), and vowel (Vowel) denote averaged attribute-group errors;
Avg. AER is computed over all attributes.}
\label{tab:cv13_aishell_aer_breakdown}
\setlength{\tabcolsep}{5pt}
\begin{tabular}{lcccc}
\toprule
\textbf{Model} & \textbf{Avg. AER $\downarrow$} & \textbf{Tone} & \textbf{Cons} & \textbf{Vowel} \\
\midrule
\multicolumn{5}{l}{\textbf{IPA-S Models}} \\
IPA-S $\times$ Tone-CAT 
& 0.0327 & 0.0488 & 0.0300 & 0.0321 \\
IPA-S $\times$ Tone-PT 
& 0.0356 & 0.0463 & 0.0315 & 0.0337 \\
\midrule
\multicolumn{5}{l}{\textbf{IPA-D Models}} \\
IPA-D $\times$ Tone-CAT 
& \textbf{0.0183} & 0.0330 & \textbf{0.0166} & \textbf{0.0169} \\
IPA-D $\times$ Tone-PT 
& 0.0199 & \textbf{0.0303} & 0.0172 & 0.0173 \\
\bottomrule
\end{tabular}
\end{table}


\begin{table*}[!t]
  \caption{Comparison of phoneme-level baselines and proposed attribute-level MDD models on the LATIC.
  The Pitch-aware RNN-T \cite{RNN-T} model is trained on AISHELL-3, while all our models are trained on CV13-CN. PER is reported for phoneme-level systems.}
  \label{tab:lat_mdd_twocol}
  \centering
  \scriptsize
  \setlength{\tabcolsep}{4.5pt}
  \begin{tabular}{ll l l ccccc cc cccc}
    \toprule
    \textbf{Train} & \textbf{Test} & \textbf{Model} & \textbf{MDD Type} &
    \textbf{FA} & \textbf{FR} & \textbf{TA} & \textbf{TR} &
    \multicolumn{2}{c}{\textbf{TR split}} &
    \textbf{PER (\%)} & \textbf{FAR (\%)} & \textbf{FRR (\%)} & \textbf{DER (\%)} \\
    \cmidrule(lr){9-10}
    & & & &
    & & & &
    \textbf{CD} & \textbf{WD} &
    & & & \\
    \midrule

    AISHELL-3 & LAT &
    Pitch-aware RNN-T \cite{RNN-T} & Phoneme &
    -- & -- & -- & -- & -- & -- &
    26.69 & 7.70 & 25.57 & 31.80\\

    \cmidrule(lr){1-14}

    \multirow{6}{*}{CV13-CN} & \multirow{6}{*}{LAT} &
    Wav2Vec2-XLSR-53 & Phoneme &
    95 & 11656 & 35257 & 858 & 566 & 292 &
    30.89 & 9.97 & \textbf{24.85} & 34.03 \\
    & &
    
    IPA-D $\times$ Tone-CAT & Phoneme &
    101 & 12360 & 34566 & 1138 & 821 & 317 &
    27.24 & \textbf{8.15} & 26.34 & 27.86 \\

    \cmidrule(lr){3-14}
    & &
    \multirow{4}{*}{IPA-D $\times$ Tone-PT} &
    Phoneme &
    108 & 12750 & 33886 & 1121 & 829 & 292 &
    27.93 & 8.79 & 27.34 & \textbf{26.05} \\
    
    & & &
    Consonant &
    30 & 3694 & 19682 & 410 & 341 & 69 &
    -- & 6.82 & 15.80 & 16.83 \\
    
    \cmidrule(lr){4-14}
    
    & & &
    Vowel &
    78 & 9056 & 14204 & 711 & 488 & 223 &
    -- & 9.89 & 38.93 & 31.36 \\
    
    & & &
    Tone &
    59 & 7857 & 8181 & 552 & 364 & 188 &
    -- & 9.66 & 48.99 & 34.06 \\
    \bottomrule
  \end{tabular}
\end{table*}

\begin{figure*}[t]
  \centering
  \includegraphics[width=\textwidth]{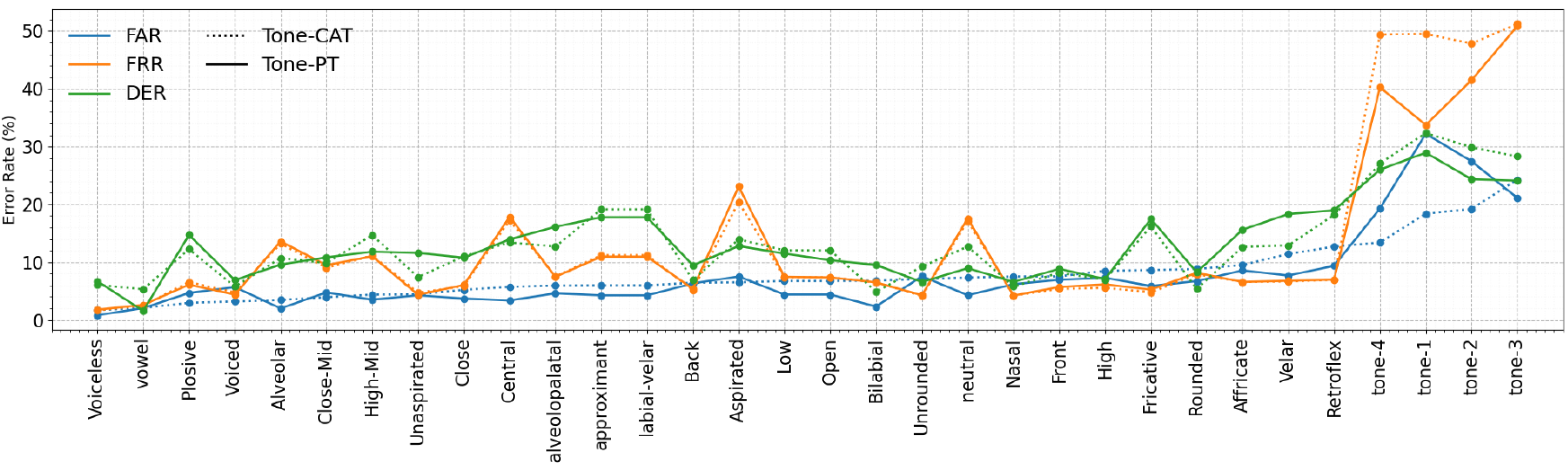}
    \caption{
    Global attribute-level MDD performance of 
    IPA-D $\times$ Tone-CAT / Tone-PT on the LATIC dataset
    sorted by FAR.
    }
  \label{fig:attribute_far}
\end{figure*}
\subsection{Mispronunciation Feedback}

\subsubsection{Phoneme-level MDD}
Table~\ref{tab:lat_mdd_twocol} reports phoneme-level MDD results on LAT, comparing our IPA-D $\times$ Tone-CAT/PT models with a Wav2Vec2-XLSR-53~\cite{w2v2} phoneme-recognition baseline and Pitch-aware RNN-T system~\cite{RNN-T} reported under a similar LAT-based setting. The attribute outputs are converted into phoneme sequences using the transcriber in Section~2. IPA-D $\times$ Tone-CAT reduces FAR from 9.97\% to 8.15\% compared with the Wav2Vec2 baseline, and achieves a lower DER than the Pitch-aware RNN-T system, reducing DER from 31.80\% to 27.86\%. These results suggest that structured phonological supervision improves overall mispronunciation detection, although the gains are metric-dependent.

Among attribute-based systems, IPA-D $\times$ Tone-PT achieves the best overall diagnostic performance, with the lowest DER (26.05\%). Compared with Tone-CAT, Tone-PT slightly increases overall FAR and FRR. In the breakdown analysis, Tone-CAT shows a lower tone FAR (8.11\%) than Tone-PT (9.66\%), suggesting that categorical tone modelling is more conservative at the phoneme level and produces fewer false alarms. The attribute-based modelling also slightly increases FRR, indicating higher sensitivity to small acoustic deviations that may trigger attribute mismatches and lead to more rejections of correct pronunciations.

\subsubsection{Attribute-level MDD}
We further analyse global attribute-level detection performance across consonant, vowel, and tone categories. The evaluation considers only attributes present in the canonical phoneme sequence. Attributes with very low occurrence frequency are excluded to avoid unstable FAR, FRR, and DER estimates. As illustrated in Figure~\ref{fig:attribute_far}, rare attributes such as \textit{dentolabial} and \textit{open-mid} are therefore omitted.

Tone-CAT and Tone-PT show negligible differences in segmental attribute detection. For tonal attributes, however, Tone-PT reduces tone FRR by about 16\% and tone DER by around 12\%, although this comes with a higher FAR. This trade-off arises because PT introduces finer-grained pitch-contour cues that improve diagnostic resolution but also increase detection sensitivity, making the system more susceptible to errors in the pitch–attribute mapping. This behaviour is consistent with the analysis in Section~3.5 and the phoneme-level MDD results in Section~3.6.1, where the \textit{offset-5} attribute shows the highest AER and contributes to weaker Tone-PT performance for Tone-1 and Tone-2 detection.

Table~\ref{tab:phoneme_attribute_far_frr_sel} reports phoneme-level and attribute-level FAR/FRR for representative confusion pairs among the top 20 substitutions in the IPA-D $\times$ Tone-PT model. Confusion pairs are extracted from substitution events only. FAR is computed on substitution tokens, while FRR is computed on correctly produced tokens of either phoneme. Attribute evaluation considers only the distinctive attributes differentiating the two phonemes. Across the top 20 pairs, each pair contains around $53 \pm 20$ tokens, reflecting sparse pronunciation errors in the dataset. With limited counts, small FA/FR or TA/TR shifts can substantially affect FAR and FRR. For example, for the consonant confusion /t/ and /\textipa{tC}/, the attributes Plosive and Alveolopalatal achieve 0\% FAR, while errors are mainly associated with the Alveolar attribute. Similarly, the vowel confusion /\textipa{i}/ and /\textipa{1}/ differs mainly in the {Front} attribute while sharing other attributes. Evaluating this attribute reduces FAR by 11.11\%, because the evaluation focuses only on distinctive features rather than penalising shared properties. For tonal confusions, attribute-level evaluation further reduces FAR by an average of $72 \pm 11$\% relative to phoneme-level detection. Overall, attribute-based MDD provides more fine-grained diagnostic insight, even with a limited number of confusion-pair occurrences.

\begin{table}[t]
\centering
\footnotesize
\setlength{\tabcolsep}{6pt}
\renewcommand{\arraystretch}{1.1}
\caption{Phoneme-level and attribute-level FAR/FRR for representative Mandarin phoneme confusions.}
\label{tab:phoneme_attribute_far_frr_sel}
\begin{tabular}{cccccc}
\toprule
 & \multicolumn{2}{c}{Phonetic} & & \multicolumn{2}{c}{Phonological} \\
\cmidrule(lr){2-3} \cmidrule(lr){5-6}
Confusion & FAR & FRR & Attribute & FAR & FRR \\
\midrule

\textipa{\:t\*s\super{h}}/\textipa{\:s} & 4.35 & 19.35 & Affricate & 1.00 & 18.60 \\
\textipa{t}/\textipa{tC} & 0.00 & 8.00 & Alveolar & 4.00 & 16.36  \\
\textipa{N}/\textipa{n} & 0.00 & 18.18 & Velar & 4.00  & 18.75  \\
\textipa{y}/\textipa{u} & 7.14 & 13.04 & Back & 0.00 & 3.77  \\
\textipa{i}/\textipa{1} & 22.22 & 6.67 & Front & 11.11 & 12.28   \\
\textipa{i}/\textipa{7} & 40.00 & 12.3 & Close-Mid & 16.67 & 12.96   \\
\textipa{7}/\textipa{a} & 16.67 & 8.70 & High-Mid & 3.70 & 12.82  \\
\textipa{W}/\textipa{i} & 16.67 & 21.74 & Unrounded & 3.45 & 8.11  \\

T2/T4 & 25.25 & 52.87 & Onset-5  & 11.59 & 34.91 \\
T4/T1 & 28.21 & 44.84 & Offset-5 & 8.56  & 30.26 \\
T3/T4 & 31.51 & 48.59 & Mid-3    & 5.70  & 51.65 \\
T2/T1 & 24.56 & 44.04 & Onset-3  & 4.43  & 48.15 \\

\bottomrule
\end{tabular}
\end{table}

\section{Conclusion}
This paper introduced a phonological-level framework for Mandarin MDD that jointly models segmental and tonal attributes using a unified wav2vec2 architecture. The proposed approach enables detailed analysis of pronunciation errors by identifying the articulatory components involved in their production. We further examined the effects of diphthong decoupling and tone pitch-contour modelling on overall system performance. Future work will focus on refining tone-attribute representations and leveraging speech synthesis to generate augmented mispronunciation data to improve MDD performance.

\newpage             
\setcounter{page}{5}    

\section{Acknowledgements}
\ifcameraready
This research includes computations on the Katana computational cluster, supported by Research Technology Services at UNSW Sydney~\cite{katana}.
\else
Anonymous.
\fi

\section{Use of Generative AI Tools}
During the preparation of this manuscript, AI was used for language editing and minor text polishing. All scientific content was developed and verified solely by the authors.

\bibliographystyle{IEEEtran}
\bibliography{mybib}
\end{document}